FUTURE RAILWAY COMMUNICATIONS

# A Survey on Future Railway Radio Communications Services: Challenges and Opportunities

Juan Moreno, José Manuel Riera, Leandro de Haro, and Carlos Rodríguez

## ABSTRACT

Radio communications is one of the most disruptive technologies in railways, enabling a huge set of value-added services that greatly improve many aspects of railways, making them more efficient, safer, and profitable. Lately, some major technologies like ERTMS for high-speed railways and CBTC for subways have made possible a reduction of headway and increased safety never before seen in this field. The railway industry is now looking at wireless communications with great interest, and this can be seen in many projects around the world. Thus, railway radio communications is again a flourishing field, with a lot of research and many things to be done. This survey article explains both opportunities and challenges to be addressed by the railway sector in order to obtain all the possible benefits of the latest radio technologies.

## INTRODUCTION

The amount of people who take a train every day in either subways, high-speed lines, or tramways is innumerable. In many of these trains, safety and non-safety services are provided through communications systems. The use of these systems is now a market trend in the industry, mostly in Europe, the United States, China, and Japan, but also in many emerging countries.

The value-added features that radio communications have provided to railways can be grouped into three types of services: safety-related, which are responsible for the safe movement of trains; operational non-safety services, including services for operators or stakeholders without safety implications, like CCTV, passenger information, remote maintenance, and sensing. Finally, the third group is devoted to providing Internet access to onboard passengers. The three of them have different exigencies due to the diverse nature of the requirements demanded of the network.

The maturity of some of these services is varied, but all of them show a lower degree of development than is desirable. For example, some operators provide Internet access for their passengers, but the available data throughputs are very poor. Moreover, the niche market condition of railways and the trend of having "one service, one radio" have driven cost steeply upward. Thus, the application of more advanced radio communication systems can bring about a lot of advantages for passengers and railway operators in all these fields: having safer and closer trains moving at higher speeds means transporting more people with better quality of service; providing customers with a good sense of security, especially in mass transit, where real-time CCTV is vital for almost every railway operator; and finally, giving a good onboard connection to the Internet implies higher incomes for operators. There are many more advantages, but those are the most relevant ones.

All these aspects are big challenges for various reasons: safety services have strong requirements in terms of reliability, availability, timing, and, of course, security. Also, operational radio telephony is a technology that belongs to the field of public safety (with all that means in terms of group calls, functional addressing, device-to-device communication, etc.). Some operational services (e.g., video streaming) are very demanding in terms of bandwidth, and providing Internet access is not a trivial task when the terminals are traveling at speeds close to 350 km/h. Finally, implementing vehicle-to-vehicle (V2V) communications could become a substantial upgrade for some services.

Today, Long Term Evolution (LTE) is a reality in public mobile communications, but it is still only a promising technology for railways. Its standardization group is trying to overcome some of the problems that made third generation (3G) technologies somewhat of a failure on railways, and as we will see, LTE is very likely to provide an excellent framework for the desired radio convergence that would be able to offer all these services over a single media, with quality of service and providing security mechanisms.

In this article we provide summarized insight into all of these challenges, mapping them to technological issues. We discuss far and near future aspects, highlighting some railway issues that are more relevant and trying to glimpse the future of the field of radio communications in railways. As far as we know, there is no survey of

*Juan Moreno and Carlos Rodríguez are with the Metro de Madrid S.A.*

*Leandro de Haro and José M. Riera are with Universidad Politécnica de Madrid (UPM).*

these characteristics in the literature.

The structure of this article is as follows. The challenges for critical services (related to safety) are summarized. Value-added operational (non-safety) services are described, and then how to provide Internet access to onboard passengersis discussed. We provide insight on some general technological aspects related to all the previous types of services; and finally, conclusions are presented.

## SAFETY SERVICES

There are two types of critical services: those related to the safety of the train itself (railway signaling) and public safety ones (including operational voice among others). In this section we mention both of them. Typically, all safety-based services need the highest safety level (SIL4 [1]), low bandwidth (less than 1 kb/s per train), significant delay constraints (less than 500–800 ms in the worst case, usually even less) and the traffic pattern is usually real-time variable bit rate (RT-VBR). Voice calls need more kilobytes per second (i.e., 64 kb/s), but it depends on the codec; a good reference for maximum jitter could be 30 ms.

### SIGNALING FOR HIGH-SPEED TRAINS

The state-of-the-art signaling system for mainline and high-speed rail is the European Rail Traffic Management System (ERTMS). In this system, both levels 2 and 3 strongly depend on the train-to-wayside communications (provided by the Global System for Mobile Communications — Railway, GSM-R, system) to work. ERTMS level 2 is a successful technology all around the world. However, level 3 (which implies the removal of track circuits; Fig. 1) shows a bit of reluctance to follow the path of level 2. The reason behind this (other than some technical issues) is the translation of costs from the infrastructure manager to the operators, because the train integrity is now guaranteed by the train itself.

ERTMS is a critical service for train safety, where trains periodically report their location, which the wayside equipment sends to train movement authorities, telling a train at each point how far it can go and how fast. This implies small-sized packets (most of the time less than 100 bytes), handovers to be completed in less than 300 ms, end-to-end delays lower than 500 ms, and connection establishment time below 8.5 s.

An industry trend for rolling stock is to assume more signaling functions at the expense of the wayside equipment, which means more importance for radio communications. As we see in the next subsection, this trend is shared by subway systems.

However, the capital expenditure (CAPEX) of an ERTMS system is quite high for some low-density traffic. Therefore, some alternatives to standard ERTMS may appear. The most relevant of them is the regional ERTMS, which divides the line into dark zones, where only one train can be inside each one. GSM-R coverage is punctual and does not require track circuits.

Carrying ERTMS data over a packet-switched network (instead of a circuit-switched one, like

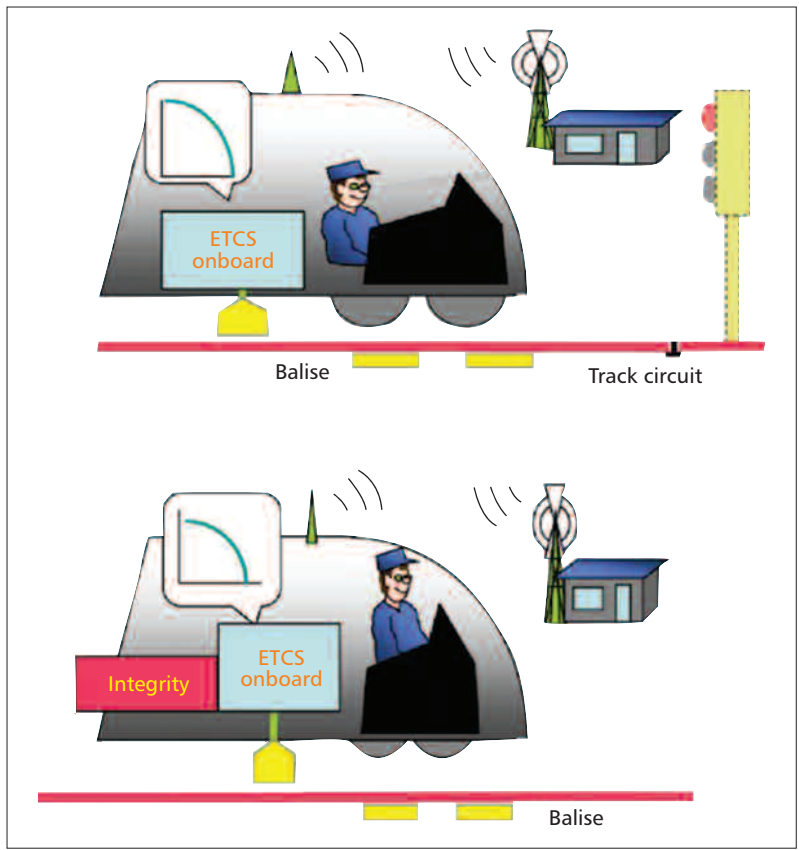

**Figure 1.** ERTMS level 2 (top) and level 3 (bottom). In level 3 there is no need for track circuits because train integrity is self-guaranteed by the train. Both levels 2 and 3 require GSM-R coverage and a network of balises.

GSM-R) is another issue that has been discussed for a while [2]. This could overcome some of GSM-R's limitations [3], extending its life a little longer.

### SIGNALING FOR SUBWAYS: CBTC

Communications-based train control (CBTC) is the ERTMS counterpart for subway trains. It is also a very successful technology, but it is not as standardized as ERTMS. CBTC systems allow trains getting closer (below 80 s headway in some cases) and safer, so it has become a de facto standard for automatic lines, driverless trains, and almost every new line.

Every vendor follows its own implementation of the radio subsystem, but they are usually based on the IEEE 802.11 family of standards. Its technical requirements are very similar to those of ERTMS, and include end-to-end delay below 800 ms and short messages (64 bytes) exchanged more or less frequently (300–500 ms). Thus, both CBTC and ERTMS demand low transmission speeds but need very reliable radio systems.

CBTC [4] technology is well established and enjoys good health, but it can be improved. A reasonable way to reduce headway (i.e., get trains closer) is to downsize delays in the whole "command" chain (processing, transmission, processing again, reaction time, etc.), but sometimes this is hard to achieve. However, the solution may be in the communications channel. If we introduce direct communications between trains (without the intervention of the interlocking or any wayside equipment), the end-to-end delay is drastically reduced.

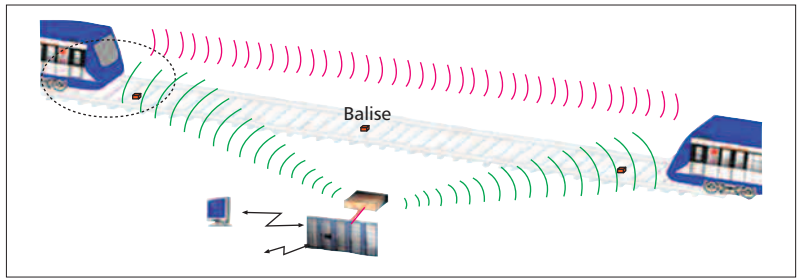

**Figure 2.** Depiction of a CBTC system with V2V capabilities.

This philosophy implies having a reliable onboard device-to-device (D2D) radio system, able to carry hard real-time information between two trains (V2V) in addition to the already available vehicle-to-infrastructure (V2I) communication (Fig. 2). This direct communication could be implemented now by using a public safety system like TETRA, or in the near future with LTE Release 13, for which direct communication capabilities will be defined. These V2V communications may be difficult to achieve if trains have obstacles between them, but if we realize that this V2V communication is meant only to get trains even closer, it makes sense. Anyway, this solution is not trouble-free, because it needs a great deal of effort in systems engineering before it comes into reality. Alstom, among other signaling vendors, is checking the feasibility of integrating this technology (called Alstom Urbalis Fluence) into its CBTC systems, but it has not released any results yet.

### TRAMWAYS: INTEGRATION ON SMART CARS PLATFORMS

The tramway is the type of railway least likely to use automation systems for operational service because it shares its way with other vehicles, like cars, buses, and even pedestrians (not only other trains). Hence, they run on sight like a car or a bus, and they can suffer accidents as those do. In the near future, smart cars and smart highways will be very common, and in this kind of service, where vehicles send information to each other to avoid crashes or improve traffic efficiency, trams could be a major actor.

Some technologies like native V2V communication standard (IEEE 802.11p [5]) or, looking at lower levels of the stack, GeoNetworking protocol (meant for ad hoc routing) should allow trams to integrate into smart car platforms, thereby improving both safety and efficiency. This is a mid-term challenge for the industry, because even the automotive part of the system is not that mature. Some of these smart car services are time-critical (typically RT-VBR), where delays should be bounded (up to 50 ms) and have low bandwidth demands.

### SIGNALING DATA OVER SATELLITES

ERTMS makes use of GSM-R and a network of beacons, but for those lines where the traffic density is not worth such an expensive deployment, there could be other alternatives, like satellites. Satellites could be used for both locating trains (aided by some GNSS systems, like GPS or Galileo in the future) and communicating with the wayside equipment. Today, two major projects are trying to validate satellite technology for railway safety services. The first one is 3InSAT, led by Ansaldo STS and the European Space Agency, now being tested in Sardinia, intended to remove track circuits, beacons, and GSM-R infrastructure using satellites for both location and communication. It also incorporates a machine-to-machine (M2M) communication system to replace the satellite signal when it is not available. The second project is SATLOC, which is very similar to 3InSAT, but only uses satellites to locate trains, whereas the communications subsystem is based on 2G/3G. The real challenge here is the setup of a safety SIL4 service over a satellite system, and also some security issues have to be addressed. The LOCOPROL/LOCOLOC project has some interesting results in this field [6].

### THE FAR FUTURE OF FREIGHT TRAINS: VIRTUAL COUPLING

A true rail safety service for the far future to be implemented over a radio communications system is virtual coupling [7]. Freight trains could circulate separated by a distance even shorter than the braking distance, because the convoy of virtually coupled trains is linked by a ultra-reliable hard real-time communications radio, and each one of them shares the same data (speed, braking commands, etc.). Thus, they behave as one single train, but the coupling between them is only virtual. Of course, it is far from becoming reality and is only a concept, but in the far future it could help jammed freight lines achieve higher capacities, because the virtual train would only occupy one slot. In some ways, this is a generalization of the V2V communication for safety introduced above.

### PUBLIC SAFETY IN RAILWAYS

Public safety (PS) communications systems are widely used all over the world for law enforcement, emergency medical services, border security, environment protection, fire fighting, search and rescue, and railways. In railways they are mostly used for operational communication between the train and the control center. They are considered a wide sense safety service because their failure does not represent a problem for the safe movement of the train, but some operators decide to interrupt service if this system is not available.

The set of functionalities required for a PS communications system basically includes the extra requirements for GSM that incorporate GSM-R (plus, obviously, the voice service). They are the following:
• Direct communication between devices (D2D)
• Group call
• Push-to-talk (fast call initiation)
• Priorization of calls, including preemption
• Data, mostly messaging
• Location-based services, like functional

addressing and location addressing
A high level of reliability and security (both authentication and integrity should be guaranteed) is also needed.

However, the two most relevant PS systems in railways (TETRA and GSM-R) have the same problem: their low capacity. TETRA is mostly used in subway environments, and GSM-R in mainlines and high-speed lines. Due to many factors (among them the support of TETRA Association) LTE is betting heavily on being the next PS communications system, but today Third Generation Partnership Project (3GPP) LTE hardly incorporates any public safety functionality. In its Releases 12 and 13 it will do so (Table 1), but this is a road paved with many difficulties, because all these PS functionalities are very challenging.

Among all these challenges, the most important of them are the following [8]:
- D2D: Node discovery, routing, radio resource management and security. It allows location-based services, and communication between nodes without a supporting infrastructure.
- Push-to-talk: PS communications need to be agile, so connection establishment time has to be very short (below 500 ms).
- Spectrum: PS channels are going to be unused most of the time, so their spectrum should be almost entirely shared, with perhaps a small part dedicated. Spectrum allocation, sharing, and management are also important issues, as always.
- Security: The degree of standardization of 3GPP LTE for PS security is very low as of 2014. This is a very relevant point because a failure here could break down the complete system.

## OPERATIONAL SERVICES

In this section, we discuss the challenges and opportunities related to operational non-safety services (typically, SIL0).

### PASSENGER INFORMATION/INFOTAINMENT

This family of services consists of providing passengers multimedia content related to the location of the train, its stops, and so on. It is a classic service, and its evolution goes toward integration with the signaling system, because it has the best knowledge of the train location. Location-based services (LBS) could provide some added value to this feature.

### CCTV

The closed circuit TV (CCTV) system is another classic operational service in railways. Due to the nature of the data (video), it is very demanding in terms of bandwidth (1 Mpx camera implies more than 1 Mb/s), and if it has a real-time basis (e.g., video streaming from the control center) it also requires bounded delays and jitter (125 ms and 25 ms, respectively). Another usual service is on-demand recording download (sometimes stored onboard). In driverless or unattended subways, it is a key pillar of the operational process.

### THE INTERNET OF THINGS

The Internet of Things (IoT) is one of the latest

| Functionality | 3GPP LTE Release | Scheduled date |
|---|---|---|
| D2D | 12 | March 2015 |
| Group call | 13 | March 2016 |
| Priorization calls | Already available | — |
| Security | 12-13 | March 2016 |
| Push-to-talk | 13 | March 2016 |
| Resilient EUTRAN | 13 | March 2016 |

**Table 1.** Public safety functionalities and the 3GPP Releases in which they are to be included.

game changers in the IT industry. This paradigm implies sensing, communicating, and aggregating all the information to obtain knowledge. Railways are more ready than other sectors to embrace it for the following reasons: modern trains have sensors almost everywhere, train-to-wayside radios are now more popular than ever, control centers are a key part of every railway network, and railway operations need to know the location of every single train. All of these are reasons for the adoption of the IoT philosophy.

However, there are some challenges and difficulties: the onboard sensors do not follow an open architecture and usually do not get out beyond the train; the railway is a very hostile environment for all this hardware; security issues should be addressed properly; the massive scale of protocols, data volume, and architectures; and finally, the conservative nature of railways does not help a new paradigm that still has many unknowns to clarify to be embraced. Anyway, if we focus on aiding operation and maintenance, it is very likely that IoT will have a lot to offer for railways.

## SERVICES FOR PASSENGERS: INTERNET ACCESS

In this section, we discuss the challenges and opportunities related to services for passengers (SIL0); usually, the philosophy here is best effort. Providing reliable and competitive Internet access to onboard passengers is not a trivial task. It has been an open problem in recent years, and due to many difficulties, it is still open. Among these complications, we find the following:
- A hostile environment, with high temperatures, vibrations, electromagnetic interference, and limited access for maintenance
- Vehicle penetration loss (VPL), usually 15–25 dB, depending on the frequency and type of vehicle
- Cyber-security
- Development of an attractive business model that offsets the high CAPEX required
- Frequent handovers
- The presence of tunnels

There are several solutions with more or less maturity [9] and some (relative) success stories,

*Relays are used to improve cell coverage, backhaul, and to increase spectral efficiency. There are many types, subtypes and classification criteria, but all mobile relays have one thing in common: they are on-board. Devices of this kind have been standardized by 3GPP LTE, WiMAX, and many others.*

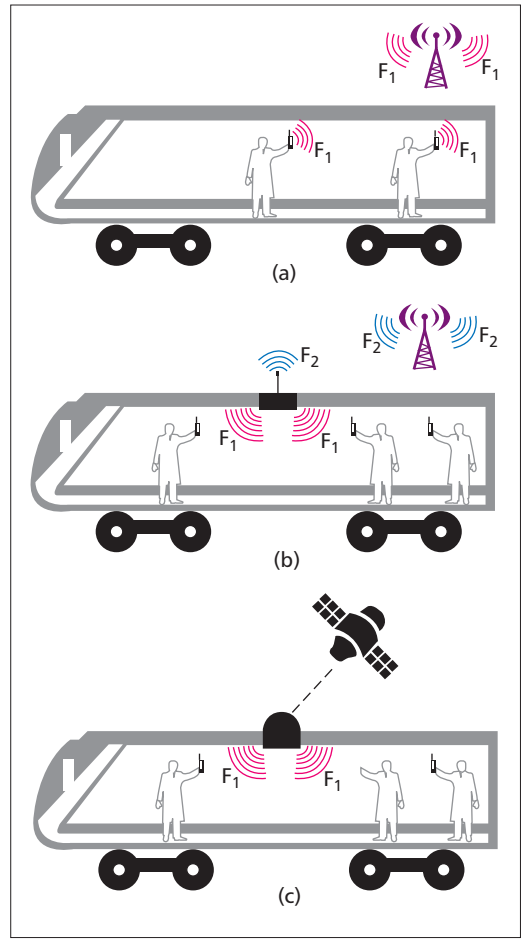

**Figure 3.** Methods for providing Internet access to on-board passengers: a) by a mobile relay and a cellular network; b) by a cellular network; and c) by a satellite system.

like that implemented by Thalys on its high-speed trains. However, they lack the desired quality that could make this service a reference and a draw for customers (in both high-speed lines and subways). Among all current and near future solutions, two stand out from the rest (Fig. 3):
• Mobile relay (MR)-based [10]
• Satellite-based

### MOBILE RELAYS

Relays are used to improve cell coverage and backhaul, and increase spectral efficiency. There are many types, subtypes, and classification criteria, but all MRs have one thing in common: they are onboard. Devices of this kind have been standardized by 3GPP LTE (starting from Release 11), WiMAX, and many others.

Its function is to split the direct link between the onboard user equipment (UE) and the base stations (BSs) into two segments. Using an MR, there are several links between the UEs and the MR, but only one between the MR and the BS. The main advantages are:
• Placing the MR antenna on the roof of the train, VPL could be overcome, allowing the use of more efficient modulation schemes and therefore higher bit rates).

• Doppler, multipath, and other undesired effects could be abridged using digital signal processing (DSP) techniques in the MR, techniques that are too expensive to be implemented in cheaper UEs.
• Group handover decreases signaling traffic.
• A toehold for operators to develop a business model.
• Larger UE battery lives.
And the major drawbacks are the following:
• Hostile environment and difficult maintenance conditions
• Increases end-to-end latency
• "All or nothing" group handover (if it fails, affects everybody)
• Need to handle interference/scheduling issues with fixed base stations
• Integration with train's systems (TCMS, etc.)

### SATELLITES

Satellites were the first solution put into practice (Thalys' case), and they have many advantages and limitations. The advantages are their simplicity (no need for a terrestrial network, so it is a good solution for railway operators that do not have an agreement with infrastructure administrators) and low CAPEX. The first and most important limitation is the need for a backup technology when trains are in tunnels, and the second is the high operational expenditure (OPEX) implied (other solutions like MR also have significant costs). To properly communicate with satellites, it is usually necessary to place huge antennas on trains' roofs, which has an impact on gauge and aerodynamics. Many research projects have explored the possibility of providing Internet access to onboard passengers [11], but most of them have failed, due to either the technical complexity or the lack of an adequate business plan.

### CYBER SECURITY

Increasingly, cyber security is a major issue on every communication service, not only in those related to providing Internet access. It is true that cyber attacks on railways have either not happened or not leaked out. The only exception (but very few details were given) is the Shenzhen Subway incident in 2012, where 3G service was shut off for a day after trains unexpectedly stopped. It is clear that security aspects cannot be ignored anymore. To investigate this type of issue, the European Project SECRET [12] arose, which intends to study all the electromagnetic risks and threats related to the railway environment. Finally, the coexistence of all these different types of services (safety-related, operational, and Internet access) is also a risk that should be addressed.

## TECHNOLOGICAL ISSUES

Here we discuss some challenges that are either not related to any of the three types of services previously discussed or related to all of them.

### RADIO CONVERGENCE

A very common scenario in these services that need train-to-wayside communications is "one service, one radio." Thus, legacy radio telephony, TETRA or GSM-R, multipurpose radios based on IEEE 802.11 b/g, signaling radios for subway trains, and so on imply that in one single train there could be as many as four different radio systems. This is due to the different nature of each one (narrow/wideband, analog/digital, critical/non-critical, trunking, IP, etc.) and the different timing for their commissioning.

However, this means very high CAPEX and OPEX, which could be avoided by using a single convergent radio, a system that aggregates all these traffic flows (Fig. 4), handling them with proper QoS and security policies. Today, the best candidate for such a convergence is LTE, above all if the public safety features explained earlier succeed in being incorporated in the 3GPP LTE standard. Moreover, betting on LTE implies IP convergence.

Another issue to take into account is the obsolescence of some of these systems (especially GSM-R, which has faced some serious limitations from its very beginning).

## WITHDRAWAL OF ONBOARD WIRING

Years ago, some operators started transmitting some car-to-car data through an IEEE 802.11g dedicated link, avoiding mechanical coupler connectors and increasing the available bandwidth. This was the kick-off for the wire-removal trend in the rolling stock scenario. On a modern train we have hundreds of meters with all kinds of wires (supply, train buses, car buses, RF cables, network, etc.) with their associated connectors. All these wires cause a lot of breakdowns and huge setup and maintenance costs, so their replacement by wireless links would be a significant improvement for railway operators. This is the idea behind one of the packages of the European R&D Project Roll2Rail, which is scheduled to be launched this year.

## HIGH-SPEED SCENARIOS

The train speed record is 574.8 km/h, and it was achieved by an Alstom train in France in 2007. However, high-speed trains usually have velocity peaks of 300–350 km/h, and Maglevs rarely run faster than 430 km/h (at least in passenger service). At such fast speeds, the Doppler effect and spreading are much more demanding than at pedestrian or car speeds, causing multipath and spreading.

Another problem could be the estimation of the channel, because if reference signals take samples at a period higher than the coherence time of the channel, system performance may decay. This channel's coherence time is inversely proportional to the Doppler shift, which depends on both carrier frequency and vehicle speed. For example, when a train travels at 300 km/h and the carrier frequency is 2.4 GHz, the Doppler shift is 4.8 kHz, and the coherence time is 88 µs. Thus, to let reference signals follow channel variations properly, we should consider pilot patterns that estimate the channel with a period no longer than 88 µs.

## CHANNEL MODELLING

Having an accurate channel model is key to deploying efficient communication systems.

| Parameter | Satellite | Mobile relay |
|---|---|---|
| High-speed performance | Very good | Good |
| Coverage | Medium | Good |
| Security | Good | Good |
| Maturity | High | Low |
| Bitrate | Medium–low | High |
| Delay | Very high | Low |
| QoS support | No | Good |
| IP | No | Yes |
| Cost | High OPEX Low CAPEX | Low OPEX Medium–low CAPEX |

**Table 2.** Comparison between mobile relays and communication satellites to provide Internet access to onboard customers.

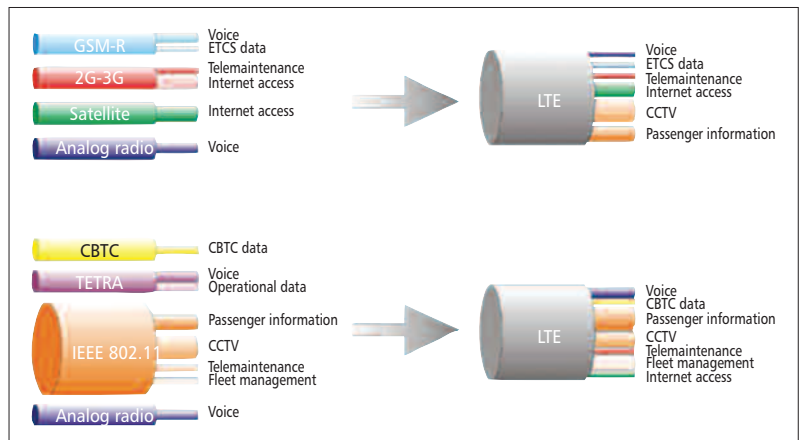

**Figure 4.** Non-convergent and convergent radio for (top) high-speed railways and (bottom) subways.

However, in V2I and V2V scenarios this is far from being taken for granted, because there are many aspects that have to be addressed. In a recent research work, Bo *et al.* [13] identify some of the pending issues and associated difficulties. Some examples are:
• In high-speed rail, the variety of scenarios (cuttings, viaducts, etc.) reaches 12 types and 18 subtypes, and most of them still have to be measured.
• In tunnels, the breakpoint that separates the near zone and far zone has not been properly calculated yet. Hrovat *et al.* [14] provide an exhaustive survey of channel modeling in tunnels.
• Train vehicle influence has to be incorporated properly to models, including VPL impact.
• In the V2V field, there is still more work to do: non-stationary channel modeling, models

> *Ee leave open a question: Is LTE the next trend in mobile communications for railways? Some aspects are favorable to this idea (3GPP LTE group has put the focus on railways and also on public safety; LTE has received some attention from the railway environment, etc.) but today it is still a promising technology, not a reality.*

that consider elevation angles, and so forth.

All these pending issues face the same difficulties to succeed: the high cost of carrying out intensive tests in railway environment.

## CONCLUSION

In this article we have summarized all the major challenges and opportunities related to radio communications that railways will meet in both the near and far future. As most usually depicted, railway services are divided into three main categories: safety-related, operational, and passenger-centric. We have outlined the most important requirements for each of them, as well as the related functionalities and the challenges behind them. Also, we take a look at other aspects like security, channel modeling, and radio convergence.

Finally, we leave open a question: Is LTE the next trend in mobile communications for railways? Some aspects are favorable to this idea (the 3GPP LTE group has put the focus on railways and also on public safety; LTE has received some attention from the railway environment, etc.) but today it is still a promising technology, not a reality (few metro lines around the world have an LTE train-to-ground radio). Of course, other options are possible, like the future 5G standard, a cognitive radio system, or even a technologically neutral one. A recent ERA report [15] provides insight on the current situation and future opportunities for operational services. Also, the recently launched European project Roll2Rail will provide some outputs in this direction.

So the big question of LTE and railways is still open. However, it is crystal clear that the future of railways will use a lot of radio-based services.

## BIOGRAPHIES

JUAN MORENO GARCÍA-LOYGORRI (juan.moreno@metromadrid.es) received his M.Sc. degree in telecommunication engineering from Universidad Carlos III de Madrid in 2006. He is also a Ph.D. candidate at Universidad Politécnica de Madrid (UPM), Spain, where he has recently completed his Ph.D. thesis, scheduled to be defended in October 2015. Since 2007, he has worked in the railway sector, first at High Speed Railways and, since October 2008 at Metro de Madrid, where he currently works in the Rolling Stock Engineering Department. He participates in many R&D projects like Roll2Rail and Tecrail.  His research interests include mobile communications, telecommunication systems in railways, propagation in tunnels, and MIMO.

JOSÉ MANUEL RIERA [M'91, SM'13] (jm.riera@upm.es) received his M.S. and Ph.D. degrees in telecommunication engineering from UPM in 1987 and 1991, respectively. Since 1993 he has been an associate professor of radio communications at UPM. His research interests are in the areas of radiowave propagation and wireless communication systems. He has directed more than 40 research projects in these fields, funded by private companies, public agencies, and national and international research programmes, including UE, ESA, and COST. He is the author of more than 130 technical papers, 100 of them published in international journals and conference proceedings or as book chapters.

LEANDRO DE HARO (leandro@gr.ssr.upm.es) received his ingeniero de telecomunicación degree in 1986 and his Doctor Ingeniero de Telecomunicación degree (Apto cum laude) in 1992, both from E.T.S.I. Telecomunicación, Departamento de Señales, Sistemas y Radiocomunicaciones, UPM. Since 1990, he has developed his professional career in the Departamento de Señales, Sistemas y Radiocomunicaciones as professor titular de universidad in the signal theory and communications area. Since 2012 he is a full professor in the Departamento de Teoría de la Señal y Comunicaciones of UPM. His research activity covers the following topics: antenna design for satellite communications (Earth stations and satellite onboard); study and design of satellite communication systems; and study and design of digital TV communication systems. He has been actively involved in several official projects and with private companies (national and international). He has also been involved in several European projects (RACE, ACTS, COST). The results of his research activity may be found in several presentations in national and international conferences as well as in published papers.

CARLOS RODRÍGUEZ SÁNCHEZ [M'06] (carlos@metromadrid.es) received his B.Sc. degree in industrial engineering (electronics and microprocessors) from the Universidad Pontificia de Comillas, Madrid, Spain, his M.Sc. degree in industrial engineering (electronics and automation) from UPM, and his Ph.D. degree in economics from the University of Alcala, Spain. He received a Ph.D. degree in electrical, electronics and system engineering from the Spanish University for Distance Education (UNED), Madrid. His industry experience includes several positions as a rolling stock and trackside engineer in the railway industry. He is currently head of the Engineering Department, Metro de Madrid S. A.. His research interests are related to railway signaling and safety in software development within industrial environments.